# A Critical Look at Ice Crystal Growth Data

Kenneth G. Libbrecht
*Norman Bridge Laboratory of Physics, California Institute of Technology 264-33, Pasadena, CA 91125*
[prepared November, 2004]

**Abstract.** I review published data relating to the growth of ice crystals from water vapor under various conditions, and I critically examine the different measurements to determine what useful information can be extracted from each. I show that most, and possibly all, of the existing growth data have been seriously distorted by systematic errors of one form or another, to varying degrees. Many data sets are dominated by systematic effects, so that the conclusions drawn from them are unreliable. I list and describe these systematic errors in some detail so that they may be avoided in future experiments. I then cautiously draw conclusions from the growth data and compare with theories pertaining to our current understanding of the crystal growth dynamics of ice.

[A better formatted and easier printed version of this paper, perhaps updated, can be found at http://www.its.caltech.edu/~atomic/publist/kglpub.htm]

## 1 Introduction

The growth of ice crystals from the vapor phase is a fascinating case study of crystal growth dynamics. The interactions of water molecules are well characterized, as is the structure of ice Ih, the normal form of ice. And yet, the growth of ice crystals from the vapor exhibits a complex behavior as a function of temperature and supersaturation that so far has not been well explained, even at a qualitative level.

One reason we do not understand ice crystal growth from vapor better is that it has proven surprisingly difficult to obtain reliable measurements of growth rates as a function of temperature and supersaturation. Since we do not understand the crystal growth dynamics of ice at a fundamental level, precise growth measurements are necessary for comparison with theoretical models of the different crystal growth mechanisms. With good measurements we can better understand the ice growth mechanisms, and we can extract numerical estimates of several important quantities that factor into the growth models. Only recently, however, have the measurements become reliable enough that one can begin to understand the growth mechanisms, and there is still plenty of room for improvement.

The purpose of this paper is to critically examine the published ice crystal growth data in order to better understand the various systematic effects that enter into the measurements. A quick look at the literature reveals that the growth data are largely inconsistent from paper to paper, so that systematic effects in the measurements have clearly been important. This task was undertaken with two principal goals in mind. First, I would like to better understand the systematic effects that were present in previous experiments, so that I and others can avoid making the same mistakes as our predecessors. Second, I would like to see just what reliable facts can extracted from the existing data, again to better guide future investigations.

It's always dangerous to criticize someone else's experiments based on the published results, since the publications invariably give only a greatly condensed version of the hardware, data, and analysis. And of course I was not present during the experiments to see just what efforts the authors expended to deal with systematic effects. I will criticize nevertheless, simply because we cannot ignore the fact that there are so many inconsistencies between the different data sets. Some of the results must be wrong, and I feel it is useful to try and understand which results were most likely dominated by systematic errors.

Clearly, the opinions expressed in this review are just that – my opinions. When I speak of systematic errors in a measurement, please bear in mind that this means *possible* systematic errors, since one cannot say for sure without an in-depth analysis of the experiment. If the authors of a paper did not discuss a particular systematic effect, and if it appears that they did not carefully consider this effect, then I am likely to conclude that this particular systematic effect *might* have been a problem in the measurements. If so, then the measurement and the results extracted from it must be considered unreliable. By unreliable I mean that while the ideas and conclusions in a paper may well be correct, they were not proven by the experiments presented. Again, we must accept that many of the published data are wrong, simply because the various inconsistencies between data sets show that they cannot all be correct.

The discussion here will focus on quantitative measurements of ice crystal growth from the vapor phase under different conditions. Many published papers provide useful qualitative results, but I will not always comment on these. Also, I did not include a number of earlier papers which investigated the ice crystal morphology diagram. The morphological changes with temperature and supersaturation are well established (although perhaps only at temperatures $T$>-30 C), so these data will not be examined here. Finally, I will not discuss crystal growth mechanisms in detail, but look mainly at growth data.

## 1.1 Notation

I will assume that ice crystal growth from the vapor can be parameterized by the Hertz-Knudsen formula [1]

$$v_n = \alpha \frac{c_{sat}}{c_{solid}} \sqrt{\frac{kT}{2\pi m}} \sigma_{surf} \qquad (1)$$

$$= \alpha v_{kin} \sigma_{surf} \qquad (2)$$

where $v_n$ is the crystal growth velocity normal to the surface, $kT$ is Boltzmann's constant times temperature, $m$ is the mass of a water molecule, $\sigma_{surf} = (c_{surf} - c_{sat})/c_{sat}$ is the supersaturation just above the growing surface, $c_{surf}$ is the water vapor number density just above the surface, $c_{sat}(T)$ is the equilibrium number density above a flat ice surface, and $c_{solid} = \rho_{ice}/m$ is the number density for ice. The parameter $\alpha$ is known as the *condensation coefficient*, and it embodies the surface physics that governs how water molecules are incorporated into the ice lattice, collectively known as the *attachment kinetics*. The attachment kinetics can be nontrivial, so in general $\alpha$ will depend on $T$, $\sigma_{surf}$, as well as the structure, chemistry, and perhaps geometry of the ice surface. If molecules striking a surface are instantly incorporated into it, then $\alpha$=1; otherwise $\alpha \leq 1$. The appearance of a crystal facet indicates that the growth of that surface is limited by attachment kinetics, so $\alpha$<1 for a facet surface. For a molecularly rough surface, or for a liquid surface, we expect $\alpha \approx 1$ [2].

Growth measurements typically measure $v_n$ as a function of $T$ and $\sigma_{surf}$, and perhaps other factors. From these measurements one hopes to extract $\alpha(T, \sigma_{surf})$ and then compare with theory to better understand the ice crystal growth dynamics. Particle and heat diffusion also affect ice crystal growth, and these are examined in some detail in the Appendix below. The results in this Appendix will be used extensively in the discussion below.

## 2 Published Data

Published papers are presented in reverse chronological order.

**"Explaining the formation of thin ice-crystal plates with structure-dependent attachment kinetics," by K. G. Libbrecht, J. Cryst. Growth 258, 168-175 (2003) [3].** This paper presented only a small amount of new ice growth data, which came from experiments similar in nature to those in [4] (see below). Diffusion was a dominant player in these data, making it effectively impossible to extract α values from growth measurements with high accuracy, as is discussed below. However, the analysis presented in [3] did not require the data to be of high accuracy, so experimental errors likely did not affect the conclusions in this paper.

**"Growth rates of the principal facets of ice between -10C and -40C," by Kenneth. G. Libbrecht, J. Cryst. Growth 247, 530-540 (2003) [5].** Ice crystals were grown on an AR-coated glass substrate in near-vacuum conditions and the growth was measured using laser interferometry. The technique was novel in that small prism crystals were first grown in air in a larger chamber before being transferred to the substrate for subsequent growth under controlled conditions. This technique avoided substrate interactions by only measuring the growth of facets not in contact with the substrate. It also allowed for the selection of crystals with good morphology. Multiple samples were grown at each temperature to see the crystal-to-crystal variations.

These measurements were the best to date in that the experiments were designed to avoid many of the systematic errors presented below in our discussion of earlier experiments. However, it is not clear that these measurements were entirely free of systematic errors. The water vapor source was a substantial distance from the growing crystals, and it was likely that numerous neighboring crystals were growing near the observed crystal, on a region of the substrate that was not observable. These effects may have been sufficient to lower the supersaturation near the observed crystals, thus distorting the data. It is difficult to estimate the magnitude of these effects, however.

These measurements showed good internal consistency, and in all cases the facet growth was consistent with nucleation-limited growth. The measured critical supersaturation increased monotonically with decreasing temperature, and it was essentially identical for the prism and basal facets.

**"Crystal growth in the presence of surface melting: supersaturation dependence of the growth of columnar ice crystals," by K. G. Libbrecht and H. Yu, J. Cryst. Growth 222, 822-831 (2001) [4].** Ice crystals were grown in free-fall in air, and the supersaturation was determined by sampling the air in the growth chamber with a hygrometer. Growth rates were determined by measuring the sizes of the largest crystals as a function of time after nucleation.

Typical crystal sizes were about 50 μm, and from Eqn. 11 in the Appendix below this implies $\alpha_{diff} \approx 0.003$. This value is sufficiently small that it is safe to assume that diffusion dominated the growth dynamics of the observed crystals (see the discussion in the Appendix). With such a small value of $\alpha_{diff}$ it becomes exceedingly difficult to extract accurate values of $\alpha$, even with careful diffusion modeling. Small systematic errors in the measurements of growth velocity or supersaturation would lead to substantial errors in the inferred values of $\alpha$, regardless of the accuracy of the modeling. Thus the conclusions in the paper are probably not very reliable.

**"Snow crystal habit changes explained by layer nucleation", by Jon Nelson and Charles Knight, J. Atmos. Sci. 55, 1452-65 (1998) [6].** Ice crystals were grown suspended from a 10-μm-OD capillary in a small chamber which contained a supercooled LiCl solution in equilibrium with water vapor above. The supersaturation $\sigma_\infty$ was determined from the concentration of the salt solution along with temperature. The measurements were done in air at one atmospheric pressure.

It appears that the growth measurements were obtained from fairly large crystals, with sizes greater than 100 μm. Thus $\alpha_{diff} \leq 0.003$ and the growth was certainly diffusion-limited to a substantial degree, implying that $\sigma_{surf} < \sigma_\infty$ during the measurements. However, no diffusion modeling was presented in the paper, so we do not have a clear picture of $\sigma_{surf}$. This problem is compounded by the fact that during the measurements some facets grew fairly rapidly while others showed no measurable growth. The rapidly growing facets will serve as a substantial water vapor sink, again reducing $\sigma_{surf}$ at nearby surfaces by some unknown amount. The paper did not report growth velocities, but rather $\sigma_\infty$ values below which the growth of free facets was too slow to be measured. Without careful modeling to determine $\sigma_{surf}$, one cannot glean much from these data. Although an intriguing study, the claims in the paper about critical supersaturations appear to be unjustified by the data presented.

One thing the paper does show, however, is that contact between an ice facet and a substrate can greatly influence the growth of the facet, at least at low growth rates. These effects may depend on substrate material, temperature, and perhaps growth rate, but clearly substrate interactions can substantially alter ice crystal growth in some cases. Thus we must assume substrate interactions can be a serious potential systematic effect in other experiments as well.

**"Temperature dependence of the growth form of negative crystal in an ice single crystal and evaporation kinetics for its surfaces," by Yoshinori Furukawa and Shigetsugu Kohata, J. Cryst. Growth 129, 571-81 (1993) [7].** The authors created "negative crystals", or voids in a single crystal, by pumping on a hollow needle embedded in ice. The sizes of the negative crystals were monitored optically as a function of time to determine evaporation rates.

The authors indicated that the void growth was typically limited by the pumping conductance of the needle, although under some conditions the growth was also influenced by the finite thermal conductivity of the ice. If the growth is conductance limited, then the determination of $\sigma_{surf}$ is not going to be very accurate. Thus the quantitative conclusions in the paper are likely not reliable.

**"The growth mechanism and the habit change of ice crystals growing from the vapor phase," by T. Sei and T. Gonda, J. Cryst. Growth 94, 697-707 (1989) [8].** Ice crystals were grown on an AR-coated glass substrate in near-vacuum conditions at various temperatures from -1 C to -30 C, as a function of supersaturation.

The data presented in this paper have low scatter and appear to be quite clean, but nevertheless there are some problems. First of all, substrate interactions were not discussed, and we see from other papers that these effects can be large [6, 9]. The authors also made repeated measurements on the same crystal by evaporating it back down after a growth cycle. Since the crystal orientation remains fixed during this process, so might the substrate interactions. Thus substrate interactions could be playing a large role and the scatter in the data would still be small.

Second, the crystals observed were quite large, up to several hundred microns, so heating may have been a problem. My experience suggests that the geometric correction *G* in Eqn. 31 in the Appendix below should be taken to be considerably less than unity to be on the safe side, perhaps $G \approx 0.2$. This indicates potential problems, especially at higher temperatures.

Third, there is some evidence that repeated measurements on the same crystal, after repeated evaporation/growth cycles, may lead to erroneous results [10]. Using the same crystal repeatedly can introduce dislocations and it also concentrates impurities on the crystal surface. This may not be a problem, but it is a concern. The low scatter in the data suggests that a single crystal was used for all the measurements at each temperature. The paper would be considerably more convincing if the authors had measured additional crystals to see how reproducible the growth velocities are. Some indication of the crystal-to-crystal variability is needed.

For all these reasons, I do not believe the data in this paper are extremely reliable, although the techniques used are an improvement over many earlier attempts.

**"Epitaxial ice crystal growth on covellite (CuS) - I. Influence of misfit strain on the growth of non-thickening crystals," by N. Cho and J. Hallett, J. Cryst. Growth 69, 317-324 (1984) [11]** and **"Epitaxial ice crystal growth on covellite (CuS) - II. Growth characteristics of basal plane steps," by N. Cho and J. Hallett, J. Cryst. Growth 69, 325-334 (1984) [12].** These back-to-back papers were the latest in a series of papers going back several decades [13, 14], in which ice crystals were epitaxially grown on substrates that closely match the molecular structure of the basal plane of ice. The results have not changed dramatically over that time, and these two papers are quite clear and well-written, so I will focus on these two as representative of the rest. Also, I will only focus on those aspects of the experiments that are related to the topic at hand – understanding the growth of ice from the vapor phase at a quantitative level.

The techniques used in these experiments are quite different from the others I have been reviewing here. The authors do not monitor the overall growth of the crystal, but rather measure the velocity of macro-steps that propagate across a smooth basal surface. Alas, many of the same systematic errors are still in effect, and it appears these rather strongly influenced the data.

First of all, the authors frequently speak of "non-thickening" crystals, which means that the flat basal faces were growing at an imperceptible rate. From Figure 2 in [12], we estimate this means a thickening of less than 0.07 μm (the thickness of the propagating step) in 70 seconds, or a velocity of <1 nm/sec. Such low growth rates mean that $\sigma_{surf}$ must have been roughly $\sigma_{surf}$<0.01 at *T* = -15 C, based on the results in [5]. Since $\sigma_\infty$ was reported to be in the range from 1 to 18 percent (Figure 7 in [12]), it appears extremely likely that diffusion was limiting the growth by reducing $\sigma_{surf}$. It appears that the observed crystals – thin, flat crystals growing epitaxially on the substrate – were typically surrounded by numerous larger, non-faceted, fast-growing crystals. This is trouble, since the non-faceted crystals act as strong sinks for water vapor and can dramatically lower the supersaturation field nearby. The authors note that at very low pressures "the rapidly growing thick ... crystals ... soon interfered with the non-thickening one to result in the sublimation of the latter." Thus the experiments could not even be performed at pressures below 25 Torr.

These considerations all suggest that diffusion was strongly affecting the growth of the macro-steps, in particular because of the non-faceted neighbor crystals. This means that $\sigma_{surf} \ll \sigma_{\infty}$, which the authors do not discuss to any degree. The measurements would have been much more convincing if the authors had managed to isolate a single faceted crystal, not surrounded by non-faceted neighbors. Then the pressure perhaps could have been reduced below 25 Torr to a situation where $\sigma_{surf} \approx \sigma_{\infty}$. Until such experiments are conducted, I have to conclude that the results in these papers are heavily influenced by systematic effects. As with the other papers reviewed here, one must exercise extreme caution when making quantitative growth measurements (of any kind) in a system where the growth is largely diffusion-limited.

Over all these decades, beginning with Mason's early investigations with epitaxially grown crystals [15], the various researchers have measured that the macro-step velocity was inversely proportional to the step height. This observation has been used to infer that the growth was dominated by the surface diffusion of molecules across the basal face and onto the step, and not by bulk diffusion through the air. This conclusion *does not follow* from the observations, however. Assume for a moment there is little or no surface diffusion, so that water molecules incorporate directly onto the step, and that the growth is limited primarily by bulk diffusion through the surrounding air. Then we again find, from a straightforward examination of the diffusion equation, that the step velocity will increase with decreasing step height. The relation will not be exactly linear, but it will likely be close enough to fit the data, which are not terribly accurate. I believe the latter explanation, that the growth is primarily limited by bulk diffusion, is more likely to be the correct one.

In a similar vein, observations have been made of the velocity of approach of two neighboring macro-steps, noting the separation at which the velocity slows. These results too have been used as evidence that the growth is being limited by surface diffusion on the basal facets (e.g. [16]). And again, however, I believe a bulk diffusion model could equally well explain the data. The latter model would be nontrivial to calculate, given the complex geometries involved. It appears that the authors of the various papers did not consider these effects, at least not to the point of doing a serious model. And again, I suspect the bulk diffusion model is the correct explanation.

How, then, does one explain the observed temperature dependence of the step velocity (Figure 8 in [12]), a result that has remained roughly unchanged for several decades? Given all the various physical effects that are present in the experiments, understanding this curve is going to be difficult. I suspect that the growth of the neighboring crystals changes with temperature in some complex, systematic way, and this in turn changes the macro-step velocity. Admittedly, this is a unsatisfying explanation of the data, which likely will not readily be accepted by all. But I strongly suspect, given the above considerations, that the step growth is being limited mainly by bulk diffusion through the surrounding gas. This suggests that the measurements are then being determined by the growth of the neighbor crystals, and not by simple considerations of surface diffusion on the basal facets. Only new measurements, using isolated crystals and with careful modeling of diffusion effects, can remove these experimental questions.

In summary, with all the epitaxial growth experiments done to date, I must conclude that the measured ice growth was predominantly limited by bulk diffusion in the surrounding gas, and thus greatly affected by the presence of rapidly growing neighbor crystals. Thus the conclusions drawn from these measurements are not reliable.

**"Rate determining processes of growth of ice crystals from the vapour phase - Part II: Investigation of surface kinetic processes," by Toshio Kuroda and Takehiko Gonda, J. Meteor. Soc. Jap. 62, 563-72 (1984) [17].** Ice crystals were grown on a substrate at -30 C in near-vacuum conditions (0.3 Torr) and in a background gas of air at 250 Torr. A crude diffusion analysis was applied to the 250-Torr data to determine $\sigma_{surf}$ and thus to extract $\alpha$ from the growth velocities.

The data look good, although it appears only a small number of crystals were observed. The results at 0.3 Torr indicated roughly isometric crystals at -30 C, and the data also showed only a weak dependence of $\alpha$ on $\sigma_{surf}$. The crystal facets apparently intersected the substrate, which may have introduced some systematic errors. Also the crystals are a bit large, suggesting heating effects may have distorted the data. We cannot reliably estimate the effects of these systematic problems, however. It would have been nice to see more data, especially at different temperatures and over a broader range of supersaturations. Overall, at 0.3 Torr the results are marginally in agreement with [5].

The data at 250 Torr lead the authors to conclude that $\alpha$ changes in the presence of air, even at constant $\sigma_{surf}$. This is an intriguing result, but I believe it is only weakly supported by the data. The basic problem is that it is very difficult to extract $\alpha$ from growth velocity data when $\alpha_{diff}$ is small. For the data presented in the paper, we have $D \approx 6 \times 10^{-5}$ m$^2$/sec (at a background pressure of 250 Torr) and $R \approx 20$ μm (or larger; this value was taken from in Figure 3 in the paper). This implies $\alpha_{diff} \approx 0.02$ or perhaps somewhat smaller (since Figure 2 in the paper suggests that larger crystals were being measured).

Given this value of $\alpha_{diff}$, consider the results shown in Figure 6 of the paper. If $\alpha=0.01$, then the growth velocity would be $v = A v_{kin} \sigma_\infty$, where $A = \alpha \alpha_{diff} / (\alpha + \alpha_{diff}) = 0.0067$. But if $\alpha=0.1$, then $A=0.0167$. The latter value is about 2.5 times larger than the former, which means a factor of ten change in $\alpha$ (from 0.01 to 0.1) produces only a factor of 2.5 change in the growth velocity. The scatter in the data shown is not this high, but it's close enough to make me worry that not enough crystals were measured.

Furthermore, the analysis in the paper was done as if there were a single crystal on the substrate, when in fact there were probably a fairly large number, the others being outside the field of view of the microscope. The authors did not discuss the possible detrimental effects of neighbor crystals, and I believe these could easily have reduced the growth rates at 250 Torr by a factor of two. If that is the case, then the data at 250 Torr would be consistent with a larger value of $\alpha$, the same as at low pressure.

Given these considerations, I believe the data only weakly support the conclusion that $\alpha$ is substantially different for crystals grown in air and in near-vacuum. This result may well be correct in the end, but one cannot say for sure from these data. It remains possible that diffusion effects alone may have produced the growth rates presented in this paper.

**"Growth rates and habits of ice crystals grown from the vapor phase," by W. Beckman, R. Lacmann, and A. Blerfreund, J. Phys. Chem. 87, 4142-4146 (1983) [9].** Ice crystals were grown on a substrate at various temperatures, both in near-vacuum conditions and in nitrogen gas at various pressures, from near-vacuum to atmospheric pressure. The crystals were monitored optically and the growth rates of the different facets were apparently recovered by modeling the

image data. For the crystals grown in nitrogen gas, a crude diffusion analysis (similar to that shown in the Appendix below) was performed to estimate $\sigma_{surf}$ and thus extract $\alpha$ from the data.

Although the paper did not give many experimental details, it described a useful trick for nucleating crystals when the steady-state supersaturation is low. A pulse of water vapor was introduced into the chamber to temporarily increase $\sigma$ and thus induce nucleation on the substrate. Since the pulse was short, the nucleated crystals quickly used up the excess water vapor so the supersaturation returned to its steady-state value. This produced small, well-formed crystals on the substrate, which subsequently could be grown at low $\sigma$ levels.

The crystals were randomly oriented on the substrate, and it was observed that substrate interactions changed the crystal growth rates. For example, the authors state that under some circumstances the growth of prism facets intersecting the substrate was *five times higher* than facets that did not intersect the substrate. Oddly, the authors did not elaborate further on this point, nor did they indicate whether facets that intersected the substrate were analyzed differently than those that did not. It would seem prudent to look only at facets that did not contact the substrate, but it appears this was not the case in the paper. It may be that some of this factor of five was due to heating effects. This would depend on the crystal sizes, which are not indicated in the paper. The clear effects of substrate interactions suggest that the quantitative growth data may not be too reliable.

The authors also reported that $\alpha$ depends strongly on the background gas pressure of nitrogen in which the crystals were grown, but again it appears the data only weakly support this conclusion. As with the Kuroda and Gonda paper above [17], we estimate that $\alpha_{diff} \leq 0.01$ at a pressure of one bar, which appears to be comparable to what they claim to have measured for $\alpha$ at the higher pressures. This again means that any small systematic error in the measured growth velocities could mimic the results shown. Thus the data do not reliably support the conclusion that $\alpha$ is lower when the background gas pressure is high.

The authors present additional data on the prism anisotropy as a function of background gas pressure (figure 4 of the paper), but there is very little discussion of this data. It is not clear how the supersaturation varied with background pressure, for example, nor is there any discussion of substrate interactions. These data are intriguing, but it's hard to know how reliable the results are.

**"Interface kinetics of the growth and evaporation of ice single crystals from the vapor phase - Part III: Measurements under partial pressures of nitrogen," by W. Beckmann, J. Cryst. Growth 58, 443-451 (1982) [10].** This paper presents some of the same results as the preceding paper [9], so the discussion is similar and I will not elaborate further here. The author notes that crystals undergoing multiple growth/evaporation cycles are more likely to develop faults that can affect their growth.

**"Interface kinetics of the growth and evaporation of ice single crystals from the vapor phase - Part II: Measurements in a pure water vapour environment," by W. Beckmann and R. Lacmann, J. Cryst. Growth 58, 433-442 (1982) [18].** This paper again presents some of the same results as in [9], so the discussion is similar. The authors note that the growth of basal faces is sometimes "inhibited" to immeasurably low rates, which is true both for facets that intersect the substrate and those that do not. The authors also note that the growth rates were largely independent of crystal orientation on the substrate, except that facets not intersecting the substrate grew more slowly. The authors also noted that not all crystals grew via the same mechanisms; some exhibited nucleation-limited growth, while others showed linear or quadratic behaviors.

**"Growth rates and growth forms of ice crystals grown from the vapor phase," by T. Gonda and T. Koike, J. Cryst. Growth 56, 259-264 (1982) [19].** Ice crystals were grown on a substrate at temperatures between -30 C and -35 C, in a background gas of air, for the purpose of examining the crystal morphology as a function of supersaturation. The results are not terribly quantitative. Substrate interactions may have been important in these measurements, again serving to influence the growth by providing a source of molecular steps. Thus even the qualitative results in this paper may not be reliable.

**"The growth of small ice crystals in gases of high and low pressures," by Takehiko Gonda, J. Meteor. Soc. Japan 54, 233-240 (1976) [20].** Ice crystals were grown in free-fall in gases of helium and argon at various pressures from 0.2 to 15 bar, and at temperatures of -7 C and -15 C. Supersaturation was provided by a fog of water droplets inside the growth chamber.

The supersaturation was probably not extremely well-determined in these experiments for a number of reasons. First, evaporative cooling can lower the droplet temperatures substantially, this effect being more pronounced at lower pressures. Second, if many crystals are growing at once they must compete for water vapor, which lowers the effective supersaturation. Third, for small clouds at low pressures the chamber walls provide a substantial perturbation of the system. The authors did not discuss these possible systematic effects in much detail, so we must assume they were present to some degree.

The qualitative results presented in the paper are almost certainly correct, in spite of the potential systematic effects. At higher gas pressures, diffusion plays a larger role and the crystal growth will develop more structure, first skeletal and then dendritic. Some of the details may be unreliable, however, owing to the possible changes in supersaturation as a function of gas pressure.

**"Studies on the ice crystals using a large cloud chamber," by A. Yamashita, Kisho Kenkyu Noto, Meteor. Soc. Japan 123, 47-97 (1974)** (in Japanese) [21], (see [22] for a description of the data in English). Ice crystals were grown in free-fall inside a cloud chamber and observed after 200 seconds of growth. With this paper one has the usual concerns with cloud chambers (see the discussion above), although it appears the author worked in a large cloud, where wall effects are reduced. Furthermore, the cloud was in air at one atmosphere of pressure, so it likely had time to equilibrate in temperature. Of course, one still must consider competition between neighboring crystals. Since the crystals grew in air, the growth was certainly predominantly diffusion limited. These data show the well-known relation between morphology and temperature, in this case at a single supersaturation near the water supersaturation level.

**"Linear growth rates of ice crystals grown from the vapor phase," by D. Lamb and W. D. Scott, J. Cryst. Growth 12, 21-31 (1972) [23].** Ice crystals were grown on a substrate at various temperatures and supersaturations, and with various background gas pressures. This appears to the be the first investigation of ice crystal growth under near-vacuum conditions. The growth rates of the prism and basal facets were measured optically.

Substrate interactions are a potential problem with these measurements, since it appears that most facets intersected the substrate. The observed crystals were quite large, so heating effects were likely to have been important. The crystals tended to show rounded tops as they became larger, with a characteristic "bread-loaf" shape that indicates that the growth on top is likely limited by heating [18]. There were also apparently a number of crystals growing simultaneously on the substrate, which means neighbor interactions may have been important. These effects are especially large when a non-faceted crystal grows near a faceted crystal, since the non-faceted crystal is a greater sink for water vapor. For growth in background pressure, diffusion effects likely dominated the growth dynamics, and were not carefully modeled. These measurements were clearly influenced by a host of unmodeled systematic effects, so we expect the results are not reliable.

## 3 Discussion

The first conclusion one can draw from the above discussion is that essentially all the experiments to date have been influenced by systematic effects to some degree, so that all the results are certainly questionable at some level. While this may seem like a heretical statement, it is supported by the simple observation that the different data sets are inconsistent with one another in many respects. One can find instances in the literature where different measurements of the same quantities produced quite different numbers.

Remarkably, after many years of attempts, we still do not possess truly reliable measurements of ice crystal growth rates as a function of temperature and supersaturation. The qualitative data, giving the well-known snow crystal morphology diagram, appear to be robust and reliable. But all the quantitative data to date are questionable, in my opinion.

One might expect that ice crystal growth measurements would not be difficult, but this is apparently not the case. Clearly the experimenter must take into consideration a number of subtle factors that can affect ice crystal growth. Before summarizing what physics we can glean from the above papers, we first list the various systematic errors that have been important in past experiments. Future workers should give considerable thought into what role these effects are playing in their measurements.

### 3.1 Systematic Errors in Ice Crystal Growth Measurements

**Diffusion.** The bottom line with diffusion is that it is nearly impossible to make an accurate measurement of $\alpha$ when $\alpha > \alpha_{diff}$. Of course one can, and should, model the effects of diffusion to determine $\sigma_{surf}$ for growing crystals, but this cannot be done with perfect precision. If $\alpha_{diff}$ is small, then any small systematic error in a measurement of crystal growth will amplify into a large systematic error in $\alpha$. Since ice crystal growth measurements are typically accurate to no better than a factor of two, one needs to use small crystals and low background pressures to obtain a reliable determination of $\alpha$.

**Neighboring Crystals.** The growth of an ice crystal can be reduced by the presence of neighboring crystals, since these all act as water vapor sinks. This may be an important consideration even in near-vacuum conditions, if one is observing the slow growth of a faceted crystal. One must note that a near-vacuum environment is not the same as a pure water vapor environment. Bulk diffusion will limit growth as long as the background gas is not all water vapor, and the particle mean free path is short compared to other relevant length scales (crystal size and/or chamber size).

If a crystal under observation is surrounded by several neighbor crystals, and these each have some nonfaceted surfaces with $\alpha \approx 1$ on those surfaces, then $\sigma$ may be greatly reduced by the neighbor crystals. The best way to see this is by considering $\sigma(x)$ in the space around the growing crystals. Diffusion determines the supersaturation field $\sigma(r)$, and the fast-growing nonfaceted crystals can be modeled by boundary conditions with $\sigma \approx 0$ at the growing surfaces. If the water vapor source is far away, then the nonfaceted crystals may substantially reduce σ near the crystal under observation, even in near-vacuum conditions. The solution to this problem is to 1) make sure there aren't too many neighbor crystals, 2) make sure the neighbor crystals are far away from the crystal under observation, and 3) keep the water vapor source close.

**Cloud Dynamics.** When growing ice crystals in a cloud of water droplets, it is a mistake to assume that the supersaturation is automatically equal to the equilibrium value for liquid water at the cloud temperature. It takes time for a cloud to reach a state of near equilibrium, and indeed such a state might never be achieved in a laboratory cloud. How close one approaches equilibrium depends on the cloud size, the droplet size distribution, the background gas pressure, and other factors.

For example, large droplets in a cloud may be too warm, if they did not have time to reach the ambient air temperature after being formed. On the other hand, droplets may be cooled below ambient if evaporative cooling draws heat out faster than interaction with the surroundings can replace it. These systematic effects, as well as the perturbation from the chamber walls, become more pronounced as the background pressure is decreased. Thus it is much more difficult to produce a near-equilibrium cloud at low pressures.

Regardless of the state of the cloud, one must also worry about competition effects. When many crystals are growing inside a cloud, they all compete for water vapor, again distorting the supersaturation field.

These effects must all be modeled to gain a more accurate estimate of σ inside the cloud. Again, these problems are much worse for small clouds and when the air pressure is reduced.

**Substrate Interactions.** A number of the papers above have described how substrate interactions can affect the growth of ice crystal facets [6, 9]. It appears the main mechanism is that the intersection of an ice facet with a substrate can be a source of molecular steps, which thus increases the growth rate if the growth is nucleation-limited. It may well be that these effects are not always important, and different substrates may behave differently. One thing is clear, however – substrate interactions can easily increase ice growth rates by factors of 2-5, so this systematic effect must be taken into account. The best approach is to restrict one's measurements to facets that do not intersect the substrate, as was done in [5].

**Evaporation/Growth Cycling.** There is some evidence [10] that evaporating and regrowing crystals affects the measured growth rates. This may be from the introduction of screw dislocations or other crystal faults as the crystal is processed, or by the concentration of impurities at the ice surface. It's not clear from the existing data how much of a problem this is, but apparently it can distort the measurements. The best strategy is to avoid a lot of cycling when taking growth data.

**Nucleation Sites.** There is ample evidence that many faceted ice surfaces contain nucleation sites that serve as a source of molecular steps. These may be screw dislocations or other crystal faults, or impurities that sit on the surface. My experience (following [18]) has been that these surfaces often cannot be easily distinguished from more perfect faceted surfaces, aside from the fact that they grow much faster at low supersaturations.

**Impurities.** There is much evidence that foreign chemicals can greatly influence ice crystal growth. Impurities on the surface may impede the growth by impeding the surface diffusion of water molecules, or they may increase the growth by providing nucleation sites. So far there is little data on how clean a given experiment has to be, so this remains a relatively unexplored problem. However, crystals grown in normal laboratory air certainly exhibit morphological changes with temperature following the morphology diagram, so apparently the impurities in normal air do not disrupt the growth by a large factor.

**Crystal to Crystal Variations.** For perhaps many reasons, observations clearly show that not all seemingly simple faceted crystals grow the same. Some surfaces contain dislocations or impurities, for example, that may strongly affect the growth dynamics. Measurements of ice crystal growth rates must sample a large number of crystals if the results are to be meaningful.

**Surface Diffusion.** Finally, one much remember that a parameterization of the growth velocity in terms of $\alpha$ makes a number of implicit assumptions about the growth dynamics. One is that surface diffusion does not transport molecules from one facet to another to a substantial degree. The Schwoebel-Ehrlich effect provides a potential barrier that inhibits surface diffusion around corners [1, 24], suggesting that the exchange of admolecules between facets may not be important. But so far there is little hard evidence either way. Thus surface diffusion remains something of a wild card in understanding ice crystal growth.

**Avoiding Systematics**. While there are obviously many pitfalls that must be avoided when measuring ice crystal growth rates, it is equally obvious that the morphology diagram is robust and reproducible. It is easy to grow hexagonal prisms in normal laboratory air that are largely plate-like or columnar, depending on temperature. A great many of these crystals also exhibit a fairly high degree of hexagonal symmetry, indicating that all six prism facets grew at essentially equal rates for the life of the crystal. We are thus confident that accurate and reproducible growth measurements can be made once these various systematic errors have been dealt with properly. To date, however, the progress toward this goal has been limited, and it appears we still do not have ice crystal growth measurements that are completely trustworthy.

## 3.2 Results from Ice Crystal Prism Growth Measurements

I attempt here to list some firm results, based on the papers listed above. There are obviously large gaps in our knowledge, and, given the state of the data, future measurements may dramatically alter our picture of the crystal growth dynamics of ice.

**The morphology diagram remains largely a mystery.** It is abundantly clear that ice crystal growth in air depends rather dramatically on temperature, in a non-monotonic fashion, as described by the morphology diagram. But to date we do not even have a satisfactory *qualitative* picture of how such large variations in morphology with temperature arise from ice crystal growth.

**The parameterization in terms of a condensation coefficient appears to be valid.** This is a weak conclusion, since the existing growth data neither support it strongly nor suggest that it is invalid. It appears that $\alpha<1$ on faceted surfaces, and $\alpha\approx1$ on rough surfaces, at least to the limited accuracy of the data. Theoretically, for an infinite surface this parameterization must be valid, but this is not the case for crystals of finite size. If surface diffusion transported molecules between facets, for example, then it may appear that $\alpha>1$, meaning that surface diffusion must be explicitly included in the growth dynamics. The Schwoebel-Ehrlich effect provides a potential barrier that inhibits surface diffusion around corners [1, 24], suggesting that transport between facets would be unlikely, and so far it appears that the growth dynamics can be treated as a local phenomenon, parameterized by $\alpha$. This picture may change with additional measurements.

**A roughening transition occurs on prism facets near the melting point.** This appears to occur at a temperature of approximately $T = -2C$, and only on the prism facets; the basal facets do not show a similar transition below the melting point [25]. Snow crystals often grow as plate-like forms with no prism facets at these warm temperatures, supporting this conclusion. Since surface melting is almost certainly present this close to the melting point, the roughening transition must be at the quasi-liquid/solid interface. Observations of ice growth from liquid water also show basal faceting without prism faceting [26, 27, 28], which likely results from the same basic roughening transition.

**Facet growth is nucleation-limited over a substantial temperature range.** This conclusion is based largely on the more recent data [5], and thus strictly valid only in the temperature range -40 C $<T<$-10 C. This also only applies to facets that do not include step sources from crystal faults or impurities, but this appears to be the norm for small crystals. More data are needed, especially at higher temperatures, to confirm this result.

**The critical supersaturation is roughly the same for the prism and basal facets, and a monotonic function of temperature over a substantial temperature range.** It was speculated for many years that substantial changes in $\sigma_{crit}$ with temperature would be responsible for the morphological changes with temperature seen in the morphology diagram. Surprisingly, it now appears that $\sigma_{crit}$ changes only monotonically with temperature, although this conclusion is again based largely on [5].

The slow increase in $\sigma_{crit}$, and thus the edge free energy of a growth island, is probably indicative of surface restructuring as a function of temperature; at higher temperatures the lattice structure becomes less rigid, allowing the step edge to be smoother with a lower edge free energy. This could be quantified by molecular dynamics simulations, which appears to be feasible with current technology.

**Growth in near-vacuum is nearly isometric.** Ice crystals grown in air often grow as thin plate-like crystals or slender columnar crystals, depending on conditions according to the morphology diagram. Crystals grown in near-vacuum are much more nearly isometric, and no experiments have produced thin plates or columns in near-vacuum conditions.

**Growth becomes more structured at higher gas pressures.** There is ample evidence for this statement, especially from [20]. This also follows from theory, since structure arises in diffusion-limited growth. The interplay of faceting and branching is such that a decrease in the diffusion constant will generally lead to an increase in morphological structure.

**A new type of instability is needed to explain the growth of thin plates and columns.** The growth of nearly isometric crystals in near-vacuum is difficult to reconcile with the observation that thin plates and needles often grow in air. Diffusion actually tends to discourage the growth of ice prisms with high aspect ratios, because in such cases the supersaturation is lowest on the fast-growing surfaces [3]. For example, for the growth of thin plates at $T = -15$ C, a solution of the diffusion problem reveals that $\alpha_{prism}/\alpha_{basal} > v_{prism}/v_{basal}$, and under some circumstances we must have $\alpha_{prism}/\alpha_{basal} \approx 100$ to produce thin ice plates [3]. Under near-vacuum conditions, however, the growth is more nearly isometric and $\alpha_{prism}/\alpha_{basal} \ll 100$.

Libbrecht [3] has suggested that a new type of growth instability is necessary to reconcile the various facts. He suggested that α may change not only with temperature and supersaturation, but also with the size of the crystal facet, such that α→1 when the facet size becomes small. This is a speculative model, without a strong theoretical foundation, so additional work is needed to understand this aspect of the growth dynamics. However, it does appear that a simple α(σ,T) function cannot explain the growth of ice crystals in air and in near-vacuum, and some form of new instability is necessary to reconcile these data [3]. The true physical nature of such an instability remains enigmatic.

## 4  Appendix – Considerations From Spherical Growth

The growth of faceted ice crystals from the vapor can be influenced by attachment kinetics at the crystal surface, particle diffusion through any background gas, and crystal heating caused by the condensing vapor. In order to weigh the relative importance of these various influences, it is instructive to consider the growth of a fictitious "faceted" spherical crystal in which we treat the growing surface as if it were atomically smooth. The spherical approximation describes the growth of isometric faceted prisms fairly well (provided all facets have the same physical properties), and we will see below it is useful in other circumstances as well. Some parts of this formalism have previously been described by Yokoyama and Kuroda [29].

### 4.1  Finite Kinetics, Without Heating

Particle transport through the gas above the crystal is described by the diffusion equation

$$\frac{\partial c}{\partial t} = D\nabla^2 c \tag{3}$$

where $c(r)$ is the particle concentration surrounding the crystal and $D$ is the diffusion constant. We will first ignore latent heat deposition, so temperature is constant throughout the system, and work in terms of the supersaturation level

$$\sigma(r) = \frac{[c(r) - c_{sat}]}{c_{sat}} \tag{4}$$

where $c_{sat}$ is the equilibrium concentration above a flat ice surface. Under essentially all realistic conditions the growth is slow enough that the diffusion equation reduces to Laplace's equation $\nabla^2 \sigma = 0$, which for the spherical case has the simple solution

$$\sigma(r) = A + \frac{B}{r} \tag{5}$$

where the constants $A$ and $B$ are determined by the boundary conditions. We will assume a sphere of radius $R$ growing inside a large sphere of radius $r_{outer} \to \infty$, so that the outer boundary condition gives $A = \sigma_\infty$.

The growth velocity is given by

$$v = \frac{c_{sat} D}{c_{solid}} \frac{d\sigma}{dr}(R) \tag{6}$$

which defines the inner boundary condition. In subsequent calculations we will ignore the Gibbs-

Thomson mechanism, in which the equilibrium vapor pressure depends on the radius of curvature of the surface. This is a valid assumption because for most faceted crystals growing from the vapor the effects of attachment kinetics are nearly always much greater than the effects of the Gibbs-Thomson mechanism [30].

We write the growth velocity in terms of the Hertz-Knudsen formula to give

$$v = \alpha v_{kin} \sigma_{surf} \tag{7}$$

where $\alpha$ is the condensation coefficient, $\sigma_{surf} = \sigma(R)$, and the kinetic velocity is

$$v_{kin} = \frac{c_{sat}}{c_{solid}} \sqrt{\frac{kT}{2\pi m}} \tag{8}$$

Combining the two expressions for $v$ gives the mixed boundary condition at the inner radius

$$\frac{d\sigma}{dr}(R) = \frac{c_{solid}}{c_{sat} D} \alpha v_{kin} \sigma(R) \tag{9}$$

and the solution of the diffusion equation gives the growth velocity

$$v = \frac{\alpha \alpha_{diff}}{\alpha + \alpha_{diff}} v_{kin} \sigma_\infty \tag{10}$$

$$= \frac{\alpha}{\alpha + \alpha_{diff}} \frac{c_{sat} D \sigma_\infty}{c_{solid} R}$$

where

$$\alpha_{diff} = \frac{c_{sat} D}{c_{solid} v_{kin} R} \tag{11}$$

$$= \frac{D}{R} \sqrt{\frac{2\pi m}{kT}}$$

In the limit $\alpha_{diff} \ll \alpha$ the growth velocity becomes $v = c_{sat} D \sigma_\infty / c_{solid} R$, which describes purely diffusion limited growth, while in the opposite limit we have $v = \alpha v_{kin} \sigma_\infty$, which is valid for kinetics limited growth. For the case of ice growing at $T = -15$ C in air we have

$$\alpha_{diff}(-15C) \approx 0.15 \left(\frac{1 \mu m}{R}\right)\left(\frac{D}{D_{air}}\right) \tag{12}$$

where $D_{air} \approx 2 \times 10^{-5}$ m$^2$/sec is the diffusion constant for water vapor in air at a pressure of one atmosphere.

By writing the growth velocity in this form it becomes apparent that one simply cannot use measurements of the growth velocity $v$ to determine $\alpha$ when $\alpha > \alpha_{diff}$. In this case the growth is mostly diffusion limited, and small errors in the determination of $v$ can result in large errors in the derived $\alpha$. To measure $\alpha$ we must either use very small crystals or reduce the background gas pressure (typically $D \sim P^{-1}$, where $P$ is the background pressure).

Put another way, consider a case where $\alpha = 0.1$ and $\alpha_{diff} = 0.01$. Diffusion modeling gives that the growth velocity is

$$v = \frac{\alpha \alpha_{diff}}{\alpha + \alpha_{diff}} v_{kin} \sigma_\infty$$

and rearranging this gives

$$\alpha = \frac{\alpha_{diff}}{(\alpha_{diff} v_{kin} \sigma_\infty / v) - 1}$$

In order to get the correct result for $\alpha$, we must measure $\alpha_{diff} v_{kin} \sigma_\infty / v = 1.1$. Even a small systematic error in our measurement of $v$ will lead to huge errors in our inferred $\alpha$. The best modern ice growth experiments still have systematic errors greater than a factor of two in $v$, which means we cannot draw any firm conclusions about $\alpha$ unless we're confident that $\alpha_{diff} > \alpha$.

## 4.2  Diffusion-Limited Growth ($\alpha_{diff} \ll \alpha$), With Heating

Diffusion-limited growth from the vapor is actually a double diffusion problem, since in principle we must consider both particle diffusion to the growing crystal and thermal diffusion to remove the heat generated by condensation at the solid/vapor interface. The spherical case can again be solved exactly in the slow-growth limit. Assuming latent heat is carried away by thermal diffusion only, the temperature distribution $T(r)$ must be a solution of Laplace's equation, which means we have

$$T(r) = T_\infty + \frac{R \Delta T}{r} \tag{13}$$

where $\Delta T = T_{surf} - T_\infty$. The heat flowing away from the growing sphere is

$$\frac{dQ}{dt} = 4 \pi \kappa R \Delta T \tag{14}$$

where $\kappa$ is the thermal conductivity of the solvent gas, equal to approximately 0.025 W m$^{-1}$ K$^{-1}$ for air.

The heat flowing out must equal the heat flowing in, which is

$$\frac{dQ}{dt} = \lambda \frac{dM}{dt}$$
$$= \lambda \rho v 4\pi R^2 \qquad (15)$$

where $\lambda$ is the latent heat for the vapor/solid transition ($\lambda_{ice} = 2.8\times10^6$ J/kg) and $\rho$ is the solid density ($\rho_{ice} = 917$ kg/m³), so we have

$$\Delta T = \frac{vR\lambda\rho}{\kappa} \qquad (16)$$

To see how this temperature change affects the growth, we have the growth velocity

$$v = \frac{c_{sat}}{c_{solid}} \frac{D}{R} \Delta\sigma \qquad (17)$$

where $\Delta\sigma = \sigma_\infty - \sigma_{surf}$, which is valid in the presence of heating. For the case of heating with fast kinetics we have

$$c(R) \approx c_{sat} + \frac{dc_{sat}}{dT} \Delta T \qquad (18)$$

where our convention is that $c_{sat}$ is evaluated at $T_\infty$. Putting all this together yields the growth velocity

$$v = \frac{D}{R} \frac{c_{sat}}{c_{solid}} \frac{\sigma_\infty}{1+\chi_0} \qquad (19)$$

where

$$\chi_0 = \frac{\eta D \lambda \rho}{\kappa} \frac{c_{sat}}{c_{solid}} \qquad (20)$$

with $\eta = dlog(c_{sat})/dT$. Typical values for these parameters are

| T(C) | $c_{sat}/c_{solid}$ | $v_{kin}(\mu m/\sec)$ | $\eta$ | $\chi_0$ |
|---|---|---|---|---|
| -40 | $0.13\times10^{-6}$ | 17 | 0.11 | 0.03 |
| -30 | $0.37\times10^{-6}$ | 49 | 0.10 | 0.08 |
| -20 | $0.96\times10^{-6}$ | 131 | 0.092 | 0.18 |
| -15 | $1.51\times10^{-6}$ | 208 | 0.088 | 0.27 |
| -10 | $2.33\times10^{-6}$ | 324 | 0.085 | 0.41 |
| -5 | $3.54\times10^{-6}$ | 496 | 0.082 | 0.59 |
| -2 | $4.51\times10^{-6}$ | 635 | 0.080 | 0.74 |

where $\chi_0$ is evaluated in air at a pressure of one atmosphere. We see that the main effect of heating on diffusion-limited growth is to scale the growth by a factor of $(1+\chi_0)^{-1}$.

If the diffusion constant is large, so that $\chi_0 \gg 1$, then the growth velocity becomes limited by heating and we have

$$v \approx \frac{\kappa}{\lambda \rho \eta} \frac{\sigma_\infty}{R} \tag{21}$$

$$\approx (100 \mu m/sec)\left(\frac{1\mu m}{R}\right)\sigma_\infty$$

## 4.3 Finite Kinetics, With Heating

For the most general case of finite kinetics in addition to heating the analysis is similar to the above, and the final result is

$$v = \frac{\alpha}{\alpha(1+\chi_0)+\alpha_{diff}} \frac{c_{sat} D \sigma_\infty}{c_{solid} R} \tag{22}$$

If the diffusion constant is large, this becomes

$$v = \frac{\alpha \alpha_{cond}}{\alpha + \alpha_{cond}} v_{kin} \sigma_\infty \tag{23}$$

where

$$\alpha_{cond} = \frac{\kappa}{\eta R \lambda \rho v_{kin}} \tag{24}$$

$$\approx 0.5\left(\frac{1\mu m}{R}\right)$$

where the latter expression is for growth in air at $T = -15$ C (note $\kappa$ is roughly independent of background pressure down to low pressures).

## 4.4 Substrate Simulation

We can also use the spherical solution to examine growth on a substrate if we assume a hemispherical crystal with the flat surface held at the substrate temperature $T_{substrate}$. Then the heat flow into the substrate is approximately

$$\frac{dQ}{dt} \approx \pi G R \kappa_{ice} \Delta T' \tag{25}$$

which gives

$$\Delta T' = \frac{2\lambda \rho v R}{G\kappa_{ice}} \qquad (26)$$

where $G \approx 1$ is a geometric correction and here $\Delta T' = T_{surf} - T_{subst}$. Since $\kappa_{ice}/\kappa_{air} \approx 100$ this reduces the temperature increase by a factor of ~100 when compared to growth in air.

The analysis is again similar to the above, and gives the growth velocity in the general case

$$v = \frac{\alpha}{\alpha(1+\chi_0') + \alpha_{diff}} \frac{c_{sat} D \sigma_0}{c_{solid} R} \qquad (27)$$

where

$$\chi_0' = \frac{2\eta D \lambda \rho}{G \kappa_{ice}} \frac{c_{sat}}{c_{solid}} \qquad (28)$$

When $D$ is large this becomes

$$v = \frac{\alpha \alpha_{cond}'}{\alpha + \alpha_{cond}'} v_{kin} \sigma_0 \qquad (29)$$

with

$$\alpha_{cond}' = \frac{G \kappa_{ice}}{2\eta R \lambda \rho v_{kin}} \qquad (30)$$

and at $T = -15$ C this gives

$$\alpha_{cond}' \approx 25 G \left( \frac{1 \mu m}{R} \right) \qquad (31)$$

Note that $v_{kin}$ increases fairly strongly with temperature, so that heating is more likely to limit the growth at higher temperatures.